\documentclass[10pt, two column, twoside]{IEEEtran}

\usepackage{amssymb}
\usepackage{amsmath}
\usepackage{mathrsfs}
\usepackage{amsmath,varwidth,array,ragged2e}
\usepackage[lined,boxed,commentsnumbered, ruled]{algorithm2e}

\usepackage{tikz}
\usepackage[latin1]{inputenc}
\usetikzlibrary{arrows}
\usetikzlibrary{shapes,arrows}
\usetikzlibrary{mindmap,trees}
\usepackage{subfigure}
\usepackage{graphicx,booktabs,multirow}
\usepackage{color}

\definecolor{colorhkust}{RGB}{20,43,140}
\definecolor{colortsinghua}{RGB}{116,52,129}
\definecolor{color1}{HTML}{D0B22B}

\newcommand{\rev}{\color{black}}

\newcommand{\tabincell}[2]{\begin{tabular}{@{}#1@{}}#2\end{tabular}}
\newcolumntype{M}{>{$}l<{$}}

\begin{document}

\title{Large-Scale Convex Optimization for Ultra-Dense Cloud-RAN}
\author{Yuanming~Shi, Jun~Zhang, and~Khaled~B.~Letaief, The Hong Kong University of Science and Technology

Bo Bai and Wei Chen, Tsinghua University}

\maketitle

\begin{abstract}
The heterogeneous cloud radio access network (Cloud-RAN) provides a revolutionary way to densify radio access networks. It enables centralized coordination and signal processing for efficient interference management and flexible network adaptation. Thus, it can resolve the main challenges for next-generation wireless networks, including higher energy efficiency and spectral efficiency, higher cost efficiency, scalable connectivity, and low latency. In this article, we shall provide an algorithmic thinking on the new design challenges for the dense heterogeneous Cloud-RAN based on convex optimization. As problem sizes scale up with the network size, we will demonstrate that it is critical to take unique structures of design problems and inherent characteristics of wireless channels into consideration, while convex optimization will serve as a powerful tool for such purposes. Network power minimization and channel state information acquisition will be used as two typical examples to demonstrate the effectiveness of convex optimization methods. We will then present a two-stage framework to solve general large-scale convex optimization problems, which is amenable to parallel implementation in the cloud data center.     
\end{abstract}

\section{Introduction}
With the dramatic increase of smart mobile devices, and diversified wireless applications  propelled by the advent of mobile social networks and Internet of Things (IoT), we are in an era of mobile data deluge. In particular, mobile data traffic has recently been doubling every year, which implies an astounding 1000 times increase by 2020! Furthermore, new wireless applications bring new service requirements. For instance, intensive data services will be needed in crowded places as stadiums and in densely populated metropolitan areas, while IoT applications call for scalable connectivity with diversified quality-of-service (QoS) requirements.              

To meet these key requirements, a paradigm shift is needed in radio access networks. In particular, network densification supported by various types of techniques, including small cells \cite{Tony_2013small},  massive MIMO \cite{Rusek_SPM2013} and millimeter-wave communications \cite{Rappaport_IEEEPrecoding2014}, has been proposed as a promising approach, which will result in a dense heterogeneous network. With improved spatial reuse and traffic dependent deployment, dense networks have the potential to boost the network capacity and provide diversified wireless services. However, there are formidable challenges in order to achieve the benefit of dense heterogeneous networks, including the magnified interference issue, the high capital expenditure (CAPEX) and operating expenditure (OPEX), and mobility management, etc. Therefore, a holistic approach to deploy and manage dense networks is required, and efficient multi-tier collaboration should be supported.

The heterogeneous cloud radio access network (Cloud-RAN) \cite{Peng_HCRAN} is a promising centralized radio access technology to address the key challenges  towards network densification by leveraging recent advances in cloud-computing technology. In particular, intra-tier and inter-tier interference  can be effectively mitigated by centralized signal processing and coordination at the cloud data center. Furthermore, with elastic network reconfiguration and adaptation, the operation efficiency of the heterogeneous Cloud-RAN can be significantly improved. For example, by adaptively switching on/off radio access points, and adjusting computing resources at the cloud data center, the network can be well adapted to spatial and temporal traffic fluctuations.  

However, the dense heterogeneous Cloud-RAN brings new design challenges, mainly due to the enlarged problem size as the design parameters and the required side information grow substantially. In this article, we will  provide a holistic viewpoint for designing dense Cloud-RAN via convex optimization. It has been well recognized that convex optimization provides an indispensable set of tools for designing wireless communication systems \cite{Palomar_2010convex}, e.g., coordinated beamforming, power control, user admission control, as well as data routing and flow control. The main reason for the success of convex optimization lies in its capability of flexible formulations, efficient globally optimal algorithms, e.g., the interior-point method, and the ability to leverage convex analysis  to explore the solution structure, e.g., the uplink-downlink duality in the multiuser beamforming problem. However, in dense Cloud-RAN, with its complex architecture, as well as the large size of optimization variables and parameters, new challenges come up.

In this paper, we will present new convex optimization methods for dense Cloud-RAN, considering three key aspects in such networks. First, a convex relaxation approach will be shown to be a powerful tool to deal with design problems with complicated variables, including both discrete and continuous variables. Such problems arise frequently in dense collaborative networks, such as the network power minimization problem. We will then consider channel state information (CSI) acquisition in dense Cloud-RAN, which is critical for centralized signal processing and resource allocation. A convex regularized optimization approach is used to exploit the channel structure for the high-dimensional channel estimation problem, while a successive convex approximation algorithm is used for designing stochastic coordinated beamforming to deal with CSI uncertainty. Major design problems in dense Cloud-RAN all involve a large number of parameters and design variables, and our third consideration is on large-scale convex optimization algorithms. A general two-stage approach is proposed, which can scale to large problem sizes and exploit the parallel computing environment in the cloud data center.

\section{Network Architecture and Research Challenges}
In this section, we shall introduce the main entities of heterogeneous Cloud-RAN, including the cloud data center, mobile hauling network and radio access network, followed by an overview of new research challenges.

\subsection{Network Architecture} 
 \begin{figure}[t]
 \centering
 \includegraphics[width=0.9\columnwidth]{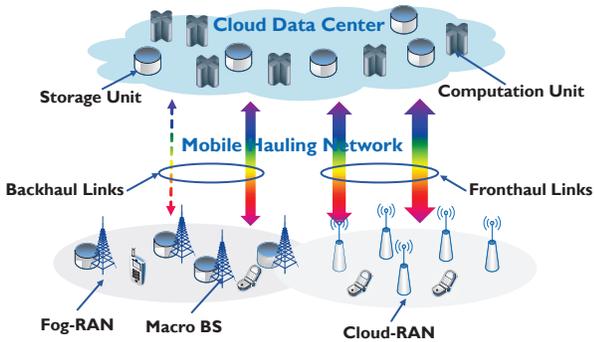}
\caption{The architecture of heterogeneous Cloud-RAN, also called as MENG-RAN. It consists of different types of radio access nodes, including Cloud-RAN and Fog-RAN.}
\label{cran}
 \end{figure}
Heterogeneous Cloud-RAN is a disruptive technology that takes advantage of recent advances in cloud-computing to revolutionize next-generation wireless networks. Its architecture is shown in Fig. {\ref{cran}}. A key feature of heterogeneous Cloud-RAN is that different radio access points will be equipped with different entities, including communication, computation, and storage units. Thus we will also call it as \emph{Multi-Entity Next Generation Radio Access Network}, or MENG-RAN. Specifically, the cloud data center serves as a central cloud infrastructure for the dense heterogeneous radio access network consisting of different types of access points. The key advantage of the heterogeneous Cloud-RAN lies in the centralized coordination at the cloud data center, supported by the mobile hauling network to transfer the information to and from different access points. In the following, we will introduce the main functionality of each entity, as well as some deployment issues.  
    
\subsubsection{Cloud Data Center}
The cloud data center consists of shared and reconfigurable computation and storage resources. Thanks to such a  shared hardware platform, both the CAPEX (e.g., via low-cost site construction) and OPEX (e.g., via centralized cooling), as well as the management effort, can be significantly reduced. Besides performing basic baseband
digital signal processing for transmission and reception,
the cloud data center can also provide cloud-computing functionalities, such as on-demand services
 via virtualization
with multiple virtual machines and resource pooling, and parallel computing
for scalable algorithm implementation.

The centralized signal processing enabled by the cloud data center is essential for the performance gains of Cloud-RAN. Specifically, with densely deployed  remote radio heads (RRHs), by applying advanced signal processing algorithms in the computationally powerful cloud data center, large-scale cooperation can be achieved, thereby improving both spectral efficiency and energy efficiency. Moreover, with centralized coordination, effective dynamic resource allocation can be provided to smooth out spatial and temporal traffic fluctuations.

\subsubsection{Radio Access Network}
In heterogeneous Cloud-RAN, access points can be divided into two categories, i.e., low-cost low-power RRHs and powerful macro BSs. 
Specifically, each
RRH only consists of a passband signal processor, an amplifier
and an A/D converter to support basic transmission and reception functionality, while the baseband signal processing is carried in the cloud data center. The data transmitted between the cloud
data center and RRHs are typically oversampled realtime I/Q digitalized baseband streams in the
order of Gbps \cite{mobile2011c}. For such access nodes, the mobile hauling network that provides high-capacity connection to the cloud data center is usually called the \emph{mobile fronthaul network}. {\rev{The typical requirements of fronthaul links are: link capacity (1-10 Gbps), latency ($\le$ 0.1 ms), and distance (1-10 km)}}. 

On the other hand, the compact macro BSs are with additional baseband signal processors and storage units. In this way, the computation and storage resources are pushed to the edge of the radio access network, which we will also call as a Fog-RAN. In this scenario, the data transmitted between the cloud data center and BSs using the packet-based interface are only side information and user messages in the order of several hundred Mbps. The mobile hauling network for such nodes is called the \emph{mobile backhaul network}, which has a low capacity requirement. {\rev{The typical requirements
of backhaul links are: link capacity (200-500 Mbps), latency ($\le$
10 ms) and distance ($\le$ 1 km)}}. By caching the content in the storage units of BSs during the idle hours, the backhaul signaling overhead can be further reduced.  However, with limited computational capability at each BS, the cooperative gain in such scenario will be lower than RRHs.

\subsubsection{Mobile Hauling Network}
A key challenge of heterogeneous Cloud-RAN is to transfer the data traffic between the cloud data center and the radio access network by the mobile hauling network. The capacity of each mobile hauling link will affect both the network performance and deployment cost, so it should be carefully picked.
 
Specifically, for the mobile fronthaul network connecting RRHs and the cloud data center, there is stringent requirement on latency and synchronization, as well as low jitter and error tolerance \cite{mobile2011c}. Both the low-cost wavelength-division multiplexing passive optical network (WDM-PON) and orthogonal frequency-division multiple access passive optical network (OFDMA-PON) are promising  candidates. With much lower capacity and latency requirements of mobile backhaul network, the  mmW (V-band frequencies, i.e., 40-75 GHz, and E-band frequencies, i.e., 71-76 GHz and 81-86 GHz) \cite{Rappaport_IEEEPrecoding2014} technology serves as a cost-effective backhaul solution.

\subsection{Research Challenges}
The new architecture of MENG-RAN will bring new opportunities, as well as new design challenges. In this paper, we shall address the main design challenges of dense Cloud-RAN from a unique perspective, i.e., we will focus on convex optimization based approaches. Convex optimization has long been recognized as a powerful tool for designing wireless networks \cite{Palomar_2010convex}. In dense Cloud-RAN, as the design problems are entering a new regime with  high-dimensional optimization variables and a large number of parameters, new design challenges come up. We will demonstrate the strength of convex optimization by developing new methodologies for the key design problems. In particular, the main focus will be on the following aspects.

\begin{itemize}
\item In dense Cloud-RAN with a large number of RRHs, it is critical to select RRHs to adapt to the temporal and spatial data dynamics, thereby improving the operating efficiency. To enable such adaptation, the new challenge comes from the composite design variables, which consist of both discrete and continuous variables for RRH selection and beamforming design, respectively. This often yields a mixed integer nonlinear programming  problem and is NP-hard.  In Section {\ref{gsbfcvx}}, with network power minimization as a representative example, we will introduce the group sparse beamforming algorithm  \cite{Yuanming_TWC2014} to efficiently solve such problems.

\item CSI is essential to various cooperation strategies of dense Cloud-RAN, but its acquisition becomes challenging as a large number of access points are involved in cooperation. To address the channel estimation challenge with limited training resources, a convex regularized optimization approach \cite{Wainwright2014structured} will be discussed in Section {\ref{csicvx}} to exploit unique structures of wireless channels.  To further deal with the resulting CSI uncertainty, a successive convex approximation algorithm is proposed to solve the corresponding highly intractable stochastic coordinated beamforming problem \cite{Yuanming_TSP14SCB}.

\item In dense Cloud-RAN, the cloud data center will typically
support hundreds of RRHs \cite{mobile2011c}, and thus all the optimization algorithms need to scale to large problem sizes. Furthermore, for many design problems, optimization algorithms should have the capability of detecting infeasibility accurately. To meet both requirements, we shall present a two-stage large-scale convex optimization framework \cite{Yuanming_Globecom2014} in Section {\ref{lscvx}} to leverage the cloud-computing environment in the cloud data center. In particular, the heterogeneous Cloud-RAN will further bring new challenges on the synchronization and distributed implementation of large-scale optimization algorithms.
\end{itemize}

\section{Convex Optimization methods for  Cloud-RAN} 
\begin{table*}[!t]
\renewcommand{\arraystretch}{1.3}
\caption{Convex Optimization methods for dense Cloud-RAN}
\label{cvx}
\centering
\begin{tabular}{l|l|l|l}
& \tabincell{l}{\bf{Group Sparse Beamforming with} \\\bf{Convex Relaxation}} & \tabincell{l}{\bf{Convex Optimization for} \\\bf{CSI Estimation and Exploitation}} & {\textbf{Large-Scale Convex Optimization}}\\
\hline
{\tabincell{c}{\bf{Problem}\\\bf{Statement}}} &  
\multirow{2}{*}{\tabincell{l}{\emph{Network Power Minimization Problem}:\\
min~~Fronthaul link power\\
~~~~~~+RRH transmit power \\
s.t.~~ Per mobile user QoS constraints\\
~~~~~~Per RRH power constraints.}} & {\tabincell{l}{\emph{High-dimensional Channel Estimation Problem}:\\
min~~Loss function + Regularizing function}} & \multirow{2}{*}{\tabincell{l}{\emph{Standard Cone Program}:\\
min~~${\bf{c}}^{T}{\boldsymbol{\nu}}$\\
s.t.~~ ${\bf{A}}{\boldsymbol{\nu}}+{\boldsymbol{\mu}}={\bf{b}}$\\
~~~~~~(${\boldsymbol{\nu}}, {\boldsymbol{\mu}}$)$\in\mathbb{R}^{n}\times {\mathcal{K}}$.\\
Problem data: ${\bf{A}}$, ${\bf{b}}$, ${\bf{c}}$.}}\\
\cline{3-3}
&  & {\tabincell{l}{\emph{Stochastic Coordinated Beamforming Problem}:\\
min~~Total transmit power\\
s. t.~~ Probabilistic QoS constraints\\
~~~~~~ Per RRH power constraints.}}& \\
\hline
{\tabincell{c}{\bf{Algorithm}\\\bf{Description}}} & 
\tabincell{l}{{\emph{Workflow:}}\\
{\textbf{Step One:}}~{\textrm{Group sparse inducing norm minimization}};\\
{\textbf{Step Two:}}~{\textrm{Fronthaul link and RRH selection}};\\
{\textbf{Step Three:}}~{\textrm{Transmit beamforming design.}}}& \tabincell{l}{{\emph{Workflow}}:\\
{\textbf{Step One:}}~{\textrm{CSI selection: determine}}\\
{``relevant" channel links;}\\
{\textbf{Step Two:}}~{\textrm{Downlink training and/or}}\\
{uplink feedback;}\\
{\textbf{Step Three:}}~{\textrm{Stochastic coordinated   }}\\
{beamforming design with mixed CSI.}} &\tabincell{l}{{\emph{Workflow:}}\\
{\textbf{Stage One:}}~{\textrm{Standard cone program}}\\
{transformation};\\
{\textbf{Stage Two:}}~{The standard form}\\{problem is solved by an }\\
{ADMM-based convex solver}.}\\
\cline{2-4}
& \tabincell{l}{\emph{Explanations}:\\
{1) In step one, the (approximated) group sparse}\\
beamforming vector is obtained by solving a convex\\
group sparse inducing norm minimization problem.\\
2) Optimal active RRHs can be obtained after step\\
two by solving a sequence of feasibility problems\\
 {\rev{to check if the remaining active RRHs can support}} \\
{\rev{the QoS requirements}}.  \\
3) In step three, optimal coordinated beamforming\\
coefficients are obtained for the active RRHs by\\
solving a convex program.} & 
\tabincell{l} {\emph{Explanations:}\\
1) A practical way to determine the optimal   \\
indices of the ``relevant" channel links \\
 is to exploit the sparsity of large-scale\\
 fading coefficients \cite{Yuanming_TSP14SCB}.\\
 2) In step two, a convex regularized  \\
 optimization method is adopted to reduce\\
  the training overhead by exploiting the\\
  channel spatial and temporal prior information.\\
 3) Only the distribution information in the\\
 mixed CSI is required to solve the stochastic\\ coordinated beamforming problem in step three.}& \tabincell{l}{\emph{Explanations:}\\
 1) Most general convex programs\\
 can be transformed into the\\
 standard cone program form with \\
 $\mathcal{K}$ as the Cartesian product of the \\
 standard cones, e.g., second-order\\
 cone. \\
 2) An operator splitting method based\\
 on ADMM algorithm is adopted to  \\
solve the standard cone programs. } \\
\hline
\end{tabular}
\end{table*} 
In this section, we shall present three design methods based on convex optimization to address the new research challenges in dense Cloud-RAN.
In particular, we will demonstrate that convex optimization has the advantage of enabling flexible formulations
and scalable algorithms, which will make it a powerful design tool for MENG-RAN. The main idea of the three proposed methods is illustrated in Table {\ref{cvx}}. {\rev{While for the illustrative purpose we mainly focus on the Cloud-RAN, the proposed methods  are generic and applicable to different coordination strategies (e.g., with CSI sharing only) among RRHs and macro BSs in heterogeneous Cloud-RAN.}}

\subsection{Group Sparse Beamforming--Convex Relaxation for Network Power Minimization}
\label{gsbfcvx}
\begin{figure}[!t]
\centering
\includegraphics[width=1\columnwidth]{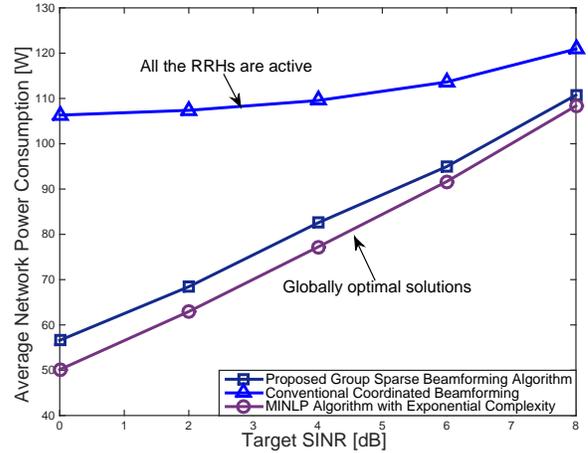}
\caption{Average network power consumption versus target signal-to-interference-plus-noise ratio (SINR). Simulation details: 10 2-antenna RRHs and 15 single-antenna mobile users uniformly and independently distributed in the square region $[-1000, 1000]\times [-1000, 1000]$ meters, the relative fronthaul link power consumption (i.e., the power saving by switching off one fronthaul link and the corresponding RRH) is set to be $(5+l) W, l=1,\dots, 10$. Standard cellular channel model is adopted \cite{Yuanming_TWC2014}. }
\label{GSBF}
\end{figure}

With densely deployed access points in heterogeneous Cloud-RAN, effective network adaptation is critical to improve the resource utilization efficiency. In particular, considering the spatial and temporal traffic fluctuation, we can dynamically select appropriate access points and mobile hauling links to serve active users. In this way, we can reduce the power consumption of both the access points and hauling links, and also reduce the signaling overhead. However, such adaptive operation will bring unique challenges for network optimization, and the design problem will need to handle both discrete (e.g., for RRH selection and user data routing) and continuous (e.g., beamforming coefficients) variables.  In this section, we will use the network power minimization problem as an example to illustrate a powerful method based on convex relaxation to deal with such design problems.

In Cloud-RAN, the network power consumption consists of two main components: the transmit power of the active RRHs, and the power consumption of their fronthaul links. By adaptively switching off some RRHs, we can save the power consumption of the corresponding fronthaul links, and thus reduce the network power. Subsequently, the network power consumption will become a composite function, i.e., the transmit power component is determined by the continuous beamforming vectors, while the fronthaul power component depends on the active RRH set, which is discrete. Thus the network power minimization problem becomes a mixed integer nonlinear programming (MINLP) problem, which is NP-hard. There are lots of other design problems in MENG-RAN that have similar structures, for which a powerful tool based on convex relaxation has recently been proposed. In \cite{Yuanming_TWC2014}, a three-step group sparse beamforming (GSBF) algorithm is proposed to minimize the network power consumption by adaptively selecting RRHs via controlling the group-sparsity structure of the aggregative beamforming vector, as illustrated in Table {\ref{cvx}}. In particular, the simulation results in Fig. {\ref{GSBF}} demonstrate that the proposed group sparse beamforming algorithm
achieves near-optimal performance. It also shows the importance of adaptively selecting RRHs to improve the energy efficiency in dense Cloud-RAN.     

The main idea of group sparse beamforming is to convexify the non-convex composite objective function via convex relaxation. In particular, the weighted $\ell_1/\ell_2$-norm of the aggregative beamforming vector of all the RRHs is shown to be the tightest convex lower bound of the original objective function. It is then used to induce the group sparsity pattern of the beamforming vector, thereby providing guidelines for RRH selection. This approach essentially exploits the group sparsity structure of the optimal solution. Specifically, all the beamforming coefficients of one RRH can be regarded as a group. When an  RRH is switched off, the corresponding beamforming coefficients in the same group will be set to be zeros simultaneously. Overall, as there will be multiple RRHs being switched off  to save the power, the optimal aggregative beamforming vector will thus have a group sparsity structure.      

For other network performance optimizations {\rev{in MENG-RAN}} that need to jointly allocate combinatorial resources (e.g., RRH selection, {\rev{data assignment}}, user scheduling and subcarrier allocation) and optimize continuous resources (e.g., beamforming coefficients, power allocation), group sparse beamforming provides a principled way to develop polynomial-time complexity algorithms. In particular, such a convex relaxation approach helps leverage the problem structures (e.g., group sparsity) via carefully deriving corresponding convex surrogates (e.g., $\ell_1/\ell_2$-norm) for the original non-convex functions (e.g., the composite objective function). Efficient convex optimization algorithms can then be applied.  Encouraging progresses have been made in applying this approach to wireless networks, {\rev{e.g., for uplink and downlink energy minimization in Cloud-RAN \cite{Rui_TWC2015GSBF} and for data assignment in backhaul links \cite{Wei_IA2014Sparse}}}. Meanwhile, there remain a variety of interesting open questions:
\begin{itemize}
\item Sparsity inducing norms for more general settings and more general utility functions should be derived, e.g., considering imperfect CSI, multicast transmission, limited fronthaul link capacity, and {\rev{the energy efficiency, which is defined as the ratio between the achievable
sum rate and the total network power consumption}}.   
\item Fast infeasibility detection
algorithms need to be developed to speedup the selection procedure (e.g., RRH selection), as a sequence of feasibility problems are needed to be solved  to make a final decision. 
\item The optimality of the group sparse beamforming algorithm should be characterized, which will be of a significant theoretical value. 
\end{itemize}

\subsection{Convex Optimization for CSI Acquisition and Exploitation}
\label{csicvx}
\begin{figure*}[t]
\centering
\subfigure[]{\includegraphics[scale=0.41]{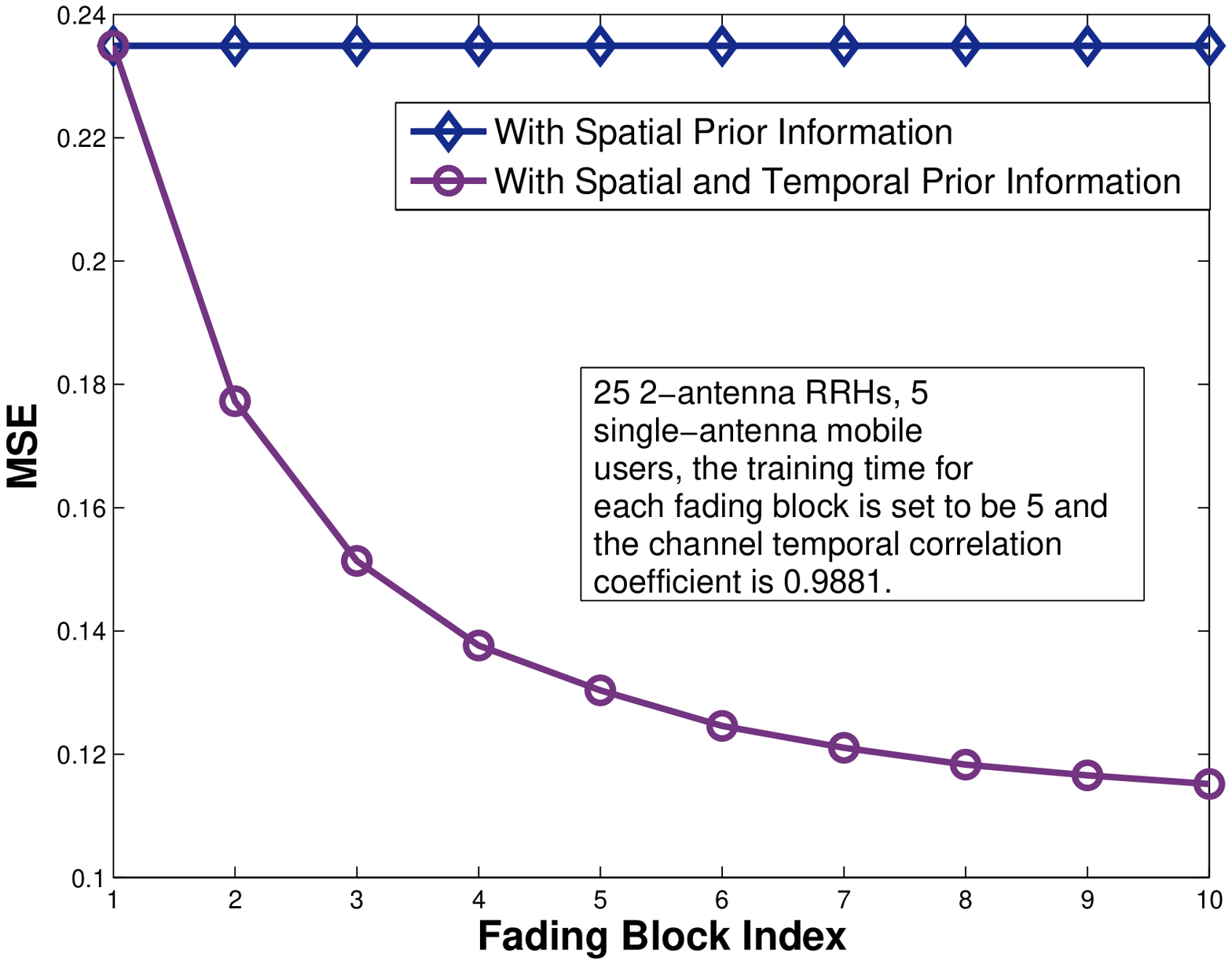}}
\subfigure[]{\includegraphics[scale=0.41]{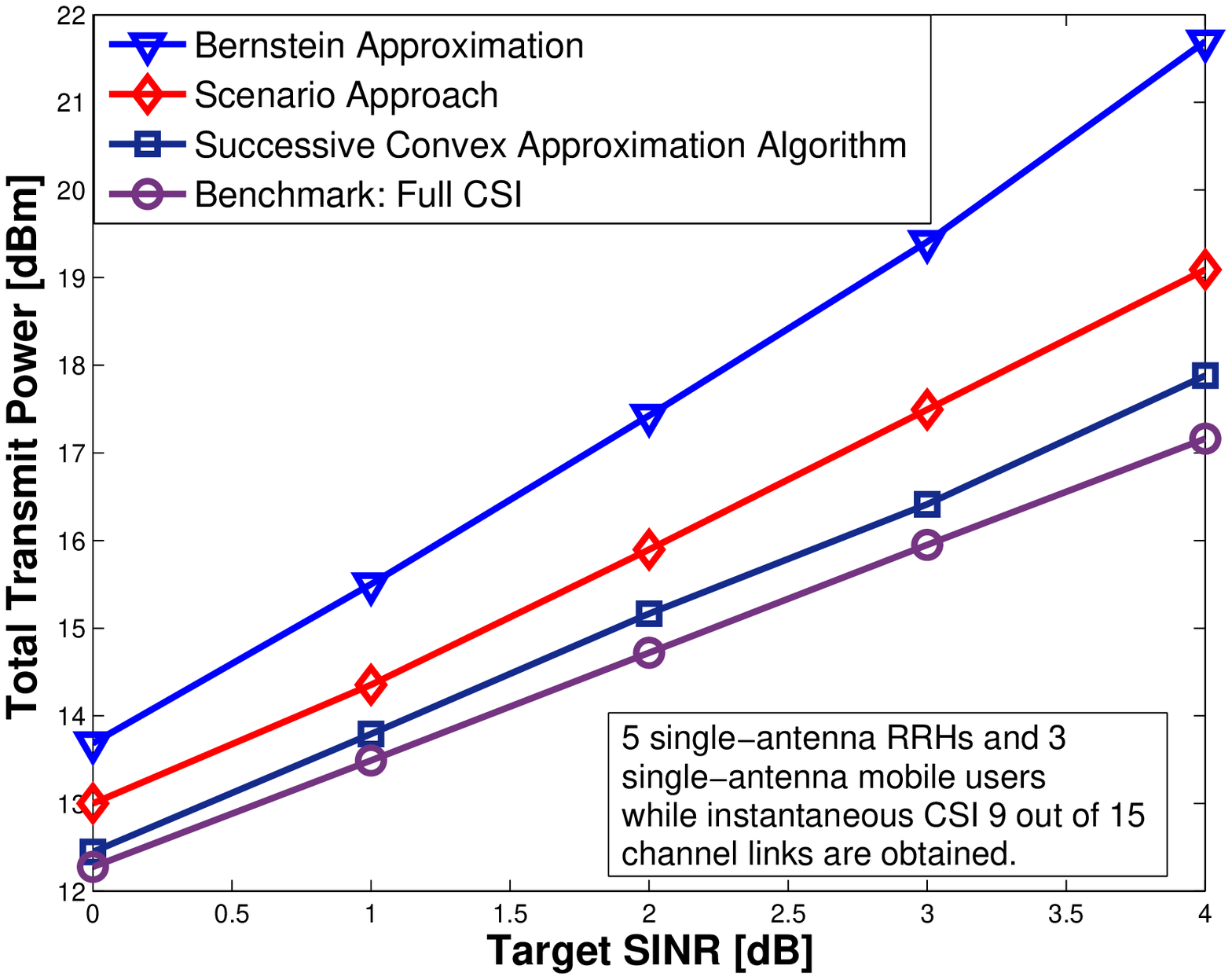}}
\caption{Convex optimization approach for CSI estimation and exploitation.  In (a), we simulate the convex regularized optimization for high-dimensional channel estimation with least-squares as the loss function and {\rev{weighted}} $\ell_1$-norm {\rev{(where the weights are defined as the inverse of the corresponding large-scale fading coefficients)}} and squared $\ell_2$-norm as the regularizing functions to exploit the spatial and temporal prior information, respectively. The performance metric is given by the mean squared error ${\rm{MSE}}=\mathbb{E}[\|\hat{\bf{H}}-{\bf{H}}\|_F^2]$.  We assume that the channel follows the first-order stationary Gauss-Markov process and we consider such a random process of length 10 blocks for each channel estimation procedure. In particular, the channel in the $i$-th block is estimated based on all the received training signals at blocks $k=1,\dots, i$.  In (b),
we simulate the stochastic coordinated beamforming problem with mixed CSI to minimize the total transmit power while satisfying  the probabilistic QoS constraints and per RRH power constraints.}
\label{csi}
\end{figure*}

CSI plays a pivotal role for effective interference management and resource allocation {\rev{as well as enabling scalable and flexible cooperation}} in heterogeneous Cloud-RAN, {\rev{e.g., network adaption in Section \ref{gsbfcvx}}}. With dense deployment of access points, CSI acquisition becomes a formidable task. In particular, due to the limited radio resources for CSI training, the training pilot length is typically smaller than the dimension of the channel. Conventional methods, such as the least square estimate, become inapplicable in such settings, and novel CSI acquisition methodologies are needed. A unique property of  Cloud-RAN with geographically distributed RRHs is the sparsity of the
large-scale fading  coefficients due to pathloss. That is, the channel links
between the RRHs and MUs that are far away will have negligible channel gains
and contribute little to system performance. A practical way to reduce the
CSI acquisition overhead in terms of training and/or feedback is to only
obtain a subset of the strongest channel  links. This is called compressive CSI acquisition \cite{Yuanming_TSP14SCB}, in which only a subset of ``relevant"
channel links {\rev{(e.g., the instantaneous channel links with the dominated large-scale fading coefficients)}} will be obtained. Furthermore,  it is critical to adjust the beamforming design so that it can effectively exploit the available channel information, while taking its uncertainty into consideration.

To address the above challenges, in this section, we shall provide a unified convex optimization based framework to develop a high-dimensional channel estimation algorithm and a successive convex approximation algorithm to deal with the  CSI uncertainty, as illustrated in Table {\ref{cvx}}.

\subsubsection{Convex Regularized Optimization for High-Dimensional Channel Estimation}
It is well known that exploiting the ``low-complexity" structures of channels can reduce the training overhead, e.g., by exploiting channel sparsity via compressed sensing. However, in dense Cloud-RAN with a large number of access points and limited radio resources, the conventional compressed sensing based approach may loss its effectiveness as it only exploits channel sparsity as a prior. This motivate us to further exploit the temporal correlation of the channels across different fading blocks to enhance the estimation performance.  

Specifically, {\rev{a convex regularized optimization approach \cite{Wainwright2014structured} can be adopted}}  to convert physical notions of channel prior information or structures (i.e., heterogenous large-scale fading and temporal correlation) into appropriate convex regularizing functions, thereby solving the underdetermined channel estimation problem with insufficient training pilots. Thus the channel estimation problem is formulated as a convex optimization problem, for which the objective consists of two parts: a loss function to measure the compatibility between the estimate and the observation, and a convex regularizing function which encodes the prior information of the channel structure. Fig. {\ref{csi}} (a) demonstrates the channel estimation performance with different available prior information. It shows that the training overhead can be reduced by exploiting the spatial and temporal prior information. This regularized approach has the potential to incorporate other structures of wireless channels for training overhead reduction. Furthermore, such convex optimization formulation also enables efficient and scalable algorithm design as will be shown in Section {\ref{lscvx}}.           

\subsubsection{Successive Convex Approximation for Stochastic Coordinated Beamforming}
With compressive CSI acquisition, the obtained CSI is actually of mixed types, consisting of a subset of imperfect instantaneous CSI  and statistical CSI for the other channel links that have not been trained.  To exploit the  available mixed  CSI, a new beamforming approach is needed to alleviate the performance degradation due to the CSI uncertainty. 
In \cite{Yuanming_TSP14SCB}, a stochastic coordinated beamforming framework based on the joint chance constrained programming  is proposed to deal with the CSI uncertainty. A probabilistic QoS  constraint is adopted, i.e., outage is allowed for each user, but its probability should be below a given threshold. Such a probabilistic constraint is motivated by the fact that most wireless networks can tolerate occasional outages. 

Although it can consider different types of CSI uncertainties, the stochastic coordinated beamforming problem is highly intractable, and thus only suboptimal algorithms exist. For instance, the scenario approach intends to  approximate the probabilistic QoS constraint by multiple ``sampling" constraints using the Monte Carlo simulation, while the  Bernstein approximation method tries to bound the chance constraint with a closed-form. {\rev{Please refer to \cite[Section IV]{Yuanming_TSP14SCB} for more details}}. But both algorithms often yield suboptimal solutions as shown in Fig. {\ref{csi}} (b). To avoid the performance loss due to suboptimal solutions, a novel successive convex approximation algorithm is proposed in \cite{Yuanming_TSP14SCB} with an optimality guarantee. It can also help us investigate the effectiveness of the proposed CSI acquisition methods. As shown in Fig. {\ref{csi}} (b), the proposed partial CSI acquisition method with stochastic coordinated beamforming can provide good performance while significantly reducing the CSI overhead. Furthermore, this figure shows that there is a tradeoff between the performance and the computational complexity, e.g., the Bernstein approximation method has the lowest computational complexity but it yields the highest transmit power.  

From the above discussion, we see that exploiting the low-complexity channel structure  becomes the ``blessing" to overcome the ``curse of dimensionality" for massive CSI acquisition in dense Cloud-RAN. In particular, convex regularized optimization provides a flexible and computationally efficient way to exploit the channel structures to reduce the training overhead. Stochastic beamforming with the convex approximation algorithm, on the other hand, provides a flexible and optimal way to deal with CSI uncertainty. {\rev{Note that the proposed CSI acquisition and exploitation method is generic and can be applied to for different coordination strategies in heterogeneous Cloud-RAN,
as it only depends on the channel properties (e.g., sparsity) for CSI acquisition
and available CSI (e.g., partial and imperfect CSI) for precoding and decoding
strategies design.}} There are, however, several interesting problems need to be further investigated.
\begin{itemize}
\item The statistical performance of the convex regularized channel estimation method should be carefully evaluated. 
\item Scalable algorithms {\rev{(e.g. the stochastic ADMM algorithm)}} need to be developed to solve the corresponding stochastic optimization problems to deal with CSI uncertainty.      
\item The fundamental performance limits of such dense distributed wireless cooperative networks with specific CSI acquisition methods and precoding/decoding strategies should be investigated.   
\end{itemize} 

\subsection{Parallelizable Large-Scale Convex Optimization Algorithms}
\label{lscvx}
\begin{table*}[!t]
\renewcommand{\arraystretch}{1.3}
\caption{Time and Solution Results with Different Convex Optimization Frameworks for the Coordinated beamforming problem to minimize the total transmit power. Details including Matlab Code can be found in \cite{Yuanming_2015largearxiv}.}
\label{mats_table}
\centering
\begin{tabular}{c|c|c|c|c|c|c}
{{Network Size ($L=K$)}} &  & 20 & 50 & 100 & 150 & 200 \\
\hline
\multirow{3}{*}{{CVX+SDPT3}}&  {{Modeling Time}} [sec] & 0.7563
&  4.4301 & N/A &
N/A & N/A \\
\cline{2-7}
 &  {{Solving Time}} [sec]& 4.2835 & 326.2513 & N/A & N/A
& N/A\\
 \cline{2-7}
 &  Objective [W]& 12.2488 & 6.5216 & N/A & N/A & N/A\\
\hline\hline
\multirow{3}{*}{{Matrix Stuffing+SCS}}&  {{Modeling Time}} [sec]& 0.0128 & 0.2401 & 2.4154 & 9.4167 &
29.5813\\
 \cline{2-7}
& {{Solving Time}} [sec]
& 0.1009 & 2.4821 & 23.8088 & 81.0023
 & 298.6224 \\
  \cline{2-7}
  &  Objective [W]& 12.2523 & 6.5193 & 3.1296 & 2.0689 & 1.5403\\
\hline
\end{tabular}
\label{timecom}
\vspace*{-10pt}
\end{table*}

\begin{figure}[!t]
\centering
\includegraphics[width=1\columnwidth]{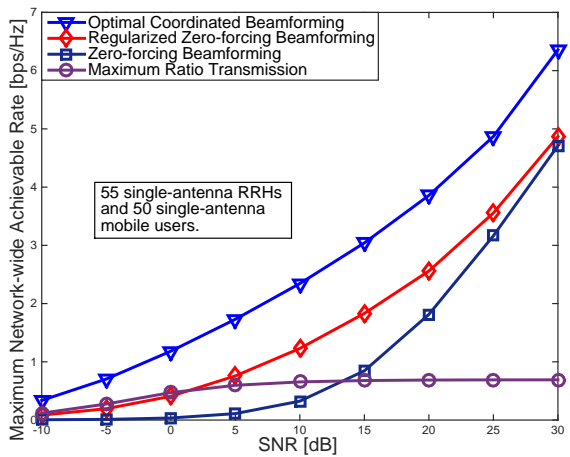}
\caption{Coordinated beamforming for max-min rate optimization versus transmit signal-to-noise ratio (SNR) in dense Cloud-RAN. The suboptimal transmission strategies only need to perform transmit power optimization as the beamforming directions are pre-fixed.}
\label{CB}
\end{figure}

We have demonstrated that convex optimization based approaches are powerful for various design problems in dense Cloud-RAN, {\rev{where a convex problem (e.g., the convex regularized optimization based channel estimation problem in Section III-B) or a sequence of convex subproblems (e.g., the three-stage GSBF algorithm in Section III-A) need to be solved}}. However, the drastically increased network density places tremendous pressure on the computational efficiency of the algorithms. For instance, for a Cloud-RAN with 100 single-antenna RRHs and 100 single-antenna MUs, the dimension of the aggregative coordinated beamforming vector (i.e., the optimization variables) will be $10^{4}$, while solving convex quadratic programs has cubic complexity using the interior-point method. Moreover,  a sequence of convex feasibility problems need to be solved for lots of design problems, e.g., for the RRH selection in the network power minimization problems. {\rev{However, most existing custom algorithms, e.g., the ADMM based algorithms \cite{Luo_2013base} and the uplink-downlink duality approach, cannot provide the certificates of infeasibility.}} Thus {\rev{effective}} feasibility detection should be embedded in the algorithm. To resolve these challenges, we provide a generic two-stage approach in \cite{Yuanming_Globecom2014}, as illustrated in Table {\ref{cvx}}. It can solve large-scale general convex optimization  problems in parallel,
with the capability of returning either the optimal solution or the certificate
of infeasibility. The time complexity of solving a general optimization problem involves two parts, i.e., the \emph{\rev{modeling time}} that transforms the problem to a standard form, and the \emph{\rev{solving time}} which solves the standard form problem. We will next demonstrate that the proposed approach can help to significantly improve the efficiency of both parts, and thus it can scale to large problem sizes.

In the first stage, the original convex program is transformed into a standard cone program with only a subspace constraint and a convex cone constraint formed by the Cartesian product of symmetric cones (e.g.,  second-order cones and positive semidefinite cones). This procedure can be done very fast using the matrix stuffing technique, which only needs to copy the problem data into the pre-stored structure of the standard cone program. As a result, all the structure information of the original convex problem is packed into the constraints. In the second stage, the structure of the standard form problem will be exploited to enable parallel computing and infeasibility detection using the alternative direction method of multipliers (ADMM) algorithm \cite{boyd2011distributed}. Such first-order method features low-or-medium-accuracy solutions, and nearly dimension-independent convergence rates. While the first stage will significant reduce the modeling time, the second stage will help reduce the solving time.    

Simulation results in Table {\ref{timecom}} demonstrate the speedups of several orders of magnitude over the state-of-the-art modeling framework CVX and interior-point solvers (e.g., SDPT3). Specifically, for the modeling time, this table shows that the proposed matrix stuffing technique can speedup about 20x to 60x compared to the parser/solver modeling framework CVX. For the solving time, taking $L=50$ as an example, the ADMM based solver SCS can speedup 130x over the inter-point solver SDPT3, which is inapplicable for large-scale problems with $L\ge 100$. Therefore, the proposed {\rev{generic}} two-stage based large-scale convex optimization framework
scales well to large-scale {\rev{performance optimization}} problems {\rev{based on the convex optimization approach in heterogenous Cloud-RAN}}. This enables us to investigate the performance of  optimal beamforming strategies in large-scale networks, which demonstrate significant gains over suboptimal transmission strategies as shown in Fig. {\ref{CB}}. This  clearly shows the importance
of developing optimal beamforming algorithms for such dense cooperative networks.   

To better exploit the advantage of the proposed approach, further efforts are needed, specified as follows. 
\begin{itemize}
\item Communication constraints and synchronization among hardware computation units should be considered to practically implement parallel algorithms.
\item Efficient and cheap subspace and cone projection (especially, for the semidefinite cone projection) algorithms are needed in the ADMM algorithm for solving large-scale standard cone programs.      
\item Other first-order methods (e.g., subgradient method
and Frank-Wolfe (projection free) method) are also worth considering to exploit the
problem structures. This, in contrast,  may be achieved by packing all the structure information into the objective function, leaving only linear constraints.

\item {\rev{To enable distributed implementation of the ADMM algorithm among macro BSs in heterogeneous Cloud-RAN, it is critical to exploit structures (e.g., decomposability \cite{Chiang_2007layering}) of the resulting problems for different coordination strategies.}}  

\end{itemize}

\section{Conclusions and Discussions}
This article discussed the benefits and design challenges of the dense heterogeneous Cloud-RAN (also called MENG-RAN). Convex optimization methods were demonstrated to be powerful tools to address the key design challenges by effectively exploiting problem structures and network properties. The presented results stand for a new paradigm for designing dense Cloud-RAN, and further investigation will be needed. Other open issues of particular interests include the optimization of the virtual machine placement and resource utilization in the virtualized cloud data center, multiterminal baseband signal compression with arbitrary topology fronthaul networks, and data flow in backhaul networks.

\bibliographystyle{ieeetr}
\bibliography{/Users/Yuanming/Dropbox/Reference/Reference}

\end{document}